\documentstyle[mprocl]{article}

\def\be{\begin{equation}}
\def\ee{\end{equation}}
\def\bea{\begin{eqnarray}}
\def\eea{\end{eqnarray}}

\begin{document}

\title{THE GEOMETRY OF DEFORMATION QUANTIZATION AND SELF-DUAL GRAVITY}
\author{Hugo Garc\'{\i}a-Compe\'an \footnote{ Present Address: {\it School
of Natural Sciences, Institute for Advanced Study, Olden Lane, Princeton
NJ 08540, USA.}} and Jerzy F. Pleba\'nski}

\address{Departamento de F\'{\i}sica \\ Centro de Investigaci\'on
y de Estudios Avanzados del IPN \\
Apdo. Postal 14-740, 07000, M\'exico D.F., M\'exico.}

\author{Maciej Przanowski}

\address{Institute of Physics \\ Technical University of \L \'od\'z\\
W\'olcza\'nska 219,  93-005 \L \'od\'z, Poland.}


\maketitle
\abstracts{A geometric formulation of the Moyal deformation for the
Self-dual Yang-Mills theory and the Chiral Model approach to
Self-dual gravity is given. We find in Fedosov's geometrical construction
of deformation quantization the natural geometrical framework associated
to the Moyal deformation of Self-dual gravity.}

\section{Introduction}

The purpose of this survey is to describe some conjectures in the geometry
of deformation quantization for Self-dual Yang-Mills (SDYM) theory and the
Chiral Model approach to Self-dual Gravity
(SDG)\cite{Hugo}\cite{Plebanone}\cite{Plebantwo} worked out in detail
there. This relation was originally suggested by I.A.B.
Strachan\cite{Straone}. He has developed a deformed differential
commutative geometry and has applied it to describe, within this
geometrical framework, the multidimensional integrable systems. Here we
intend to consider the application of some {\it non-commutative} geometry
(Fedosov's geometry \cite{Fedosov}, \cite{Wein}) to self-dual gravity. The
relation, for instance, between SDYM theory, Conformal Field Theory and
Principal Chiral Model, all them with gauge group SDiff $(\Sigma)$
(area-preserving diffeomorphism group of two-dimensional simply connected
and symplectic manifold $\Sigma$), has been quite studied only at the
algebraic level. The standard approach consists in considering a classical
field theory invariant under some symmetry group, for instance, SU$(N)$
and then take its large-$N$ limit.  In the case of Yang-Mills theory (both
full and SD) its large-$N$ limit ($N \to \infty$) is somewhat mysterious,
however it is very necessary to understand it in the searching for new
faces of integrability. Drastic simplifications in some classical
equations seem to confirm these speculations \cite{Witten}. However,
geometric and topological aspects of the correspondence SDYM and SDG
remain to be clarified.


\section{Fedosov's Geometry  and the Moyal Deformation of
Self-dual Gravity}

First of all the Principal Chiral equations can be normally written as

$$ F = dA + A\wedge A = 0, \ \ \ \ \ d \star A = 0 \eqno(1)$$ where
$\star$ is the standard Hogde operator and $A \in {\cal E} \big({\cal
G}\otimes \Lambda^1\big)$ is the connection one form. The corresponding
equations are

$$ A = g^{-1} dg, \ \ \ \ \ d \star (g^{-1}dg) = 0. \eqno(2)$$ The first
equation is the condition of {\it flat connection} and the second one is
the equation of motion. In coordinates $(x,y) \in \Omega$, $A = A_x dx +
A_y dy,$ with $ A_{\mu}(x,y)= \sum_{a=0}^{dim \ G} A^a_{\mu}(x,y) \tau_a
\in {\cal G} \otimes C^{\infty}(\Omega),$ $\mu=x,y.$Now we generalize this
gauge connection from ${\cal G}$-valued connection one-form to the
corresponding ${\cal E}(\tilde{\cal W}_D)$-valued connection one-form $
\tilde{A} = \tilde{A}_x dx + \tilde{A}_y dy,$ with ${\cal E}(\tilde{\cal
W}_D)$ the space of sections of the Weyl algebra bundle $\tilde{\cal W}$
\cite{Fedosov}

$$ \tilde{A}_{\mu}= \tilde{A}_{\mu}(x,y,p,q;\hbar)= a_{\mu} + \partial_i
a_{\mu} y^i + {1\over 6} \partial_i \partial_j \partial_k a_{\mu} y^i y^j
y^k$$
$$ - {1\over 24} R_{ijkl}\omega^{lm} \partial_m a_{\mu} y^i y^j y^k +
\cdots, \eqno(3)$$
for the case of non-flat phase-space ${\cal M}$ (with $y^1\equiv
p$, $y^2 \equiv q)$. While that for the flat case we have
$\tilde{A}_{\mu}(z,\bar{z}) = \sum_{k =0}^{\infty} {1\over k!}
\big(\partial_{i_1} \partial_{i_2} \cdots \partial_{i_k} a_{\mu} \big)
y^{i_1} y^{i_2} \cdots y^{i_k}.$ In the above formulas $
a_{\mu} = a_{\mu}(x,y,p,q;\hbar).$ The mentioned correspondence also
implies that Eqs. (1) have a counterpart in terms of Fedosov's geometry
\cite{Hugo}

$$ \tilde{F} = d\tilde{A} + \tilde{A} \buildrel{\bullet}\over {\wedge}
\tilde{A} = 0, \ \ \ \ \ d \star \tilde{A} = 0, \eqno(4) $$ where
$\tilde{A} \in {\cal E}({\cal E}(\tilde{\cal W}_D) \otimes \Lambda ^1)$
and $\tilde{F} \in {\cal E}({\cal E}(\tilde{\cal W}_D)\otimes \Lambda^2)$
and where the multiplication $\buildrel{\bullet}\over{\wedge}$ is defined
by $a \buildrel{\bullet}\over{\wedge} b = a_{[j_1\ldots j_p} \bullet
b_{l_1 \ldots l_q]} dx^{j_1} \wedge \ldots \wedge dx^{j_p}\wedge dx^{l_1}
\wedge \ldots \wedge dx^{l_q},$ for all $a = \sum_k \hbar^k a_{k,j_1
\ldots j_p}(x,y) dx^{j_1}\wedge \ldots \wedge dx^{j_p} \in {\cal
E}\big(\tilde{\cal W} \otimes \Lambda^p\big)$ and $b = \sum_k \hbar^k
b_{k, l_1 \ldots l_q}(x,y) dx^{l_1}\wedge \ldots \wedge dx^{l_q} \in {\cal
E}\big ( \tilde{\cal W} \otimes \Lambda^q \big)$. $a
\buildrel{\bullet}\over{\wedge} b$ is defined by the usual wedge product
on ${\cal M}$ and the product $\bullet$ in the Weyl algebra, $ a \bullet b
\equiv {\rm exp} \big( +{i\hbar\over 2} \omega^{ij} {\partial \over
\partial y^i}{\partial \over \partial z^i}\big) a (y ,\hbar) b (z, \hbar)
|_{z=y} = \sum^{\infty}_{k=0} (+ {i\hbar\over 2})^k {1 \over k!}
\omega^{i_1 j_1} \ldots \omega^{i_kj_k} {\partial^k a\over \partial
y^{i_1} \ldots \partial y^{i_k}} {\partial ^k b \over \partial y^{j_1}
\ldots \partial y^{j_k}}.$ The corresponding to Eqs. (2) are

$$\tilde{A} = g^{- \buildrel{\bullet}\over{1}} \bullet dg, \ \
\ \ \ d\star(g^{- \buildrel{\bullet}\over{1}} \bullet dg)= 0, \eqno(5)$$

Now we will show that Eqs. (5) can be obtained from a variational
principle from a Lagrangian of the standard ${\bf G}$ principal chiral
model. First we recall that the action which gives Eqs. (2) reads $ S =
\int_{\Omega} {\cal L} $ where $ {\cal L} ={1 \over 2} {\rm Tr} ( g^{-1}
dg \wedge \star g^{-1}dg),$ where $g: \Omega \to {\bf G}$ and $d$ is the
exterior differential on $\Omega$ {\it i.e.} $d = dx \partial_x + dy
\partial_y$ and Tr is an invariant form on the Lie algebra of {\bf G},
$Lie({\bf G}) = {\cal G}$. Here we have assumed that {\bf G} is
semisimple. The above action can be generalized to Fedosov's geometry as
follows $ S^{\bullet} = \int_{\Omega} {\cal L}^{\bullet}$ where $ {\cal
L}^{\bullet}= - { \hbar^2 \over 2}tr\big( \tilde{A}
\buildrel{\bullet}\over{\wedge} \star \tilde{A} \big) = - {\hbar^2 \over
2} tr \big(g^{- \buildrel{\bullet}\over{1}} \bullet dg
\buildrel{\bullet}\over{\wedge} \star g^{-\buildrel{\bullet}\over{1}}
\bullet dg \big).$ where `$tr$' is the Fedosov's trace
\cite{Fedosov}\footnote{ Fedosov's trace $`tr'$ is defined for flat phase
space as $tr(a)  := \int_{{\bf R}^{2n}} \sigma(a) {\omega^n \over n!}$
with $a\in {\cal E}(\tilde{\cal W}_D)$ and $\sigma: {\cal E}(\tilde{\cal
W}_D)  \to {\cal Z}$ is a bijection.} and $g: \Omega \to {\bf
G}_{\bullet}$ is the generalized gravitational uniton \cite{Plebantwo}. In
the case of flat phase-space $R_{ijkl} = 0$, the trace can be expressed by

$$ {\cal L}^{\bullet} = - {\hbar^2 \over 2}
tr(\tilde{A}\buildrel{\bullet}\over{\wedge} \tilde{A})= - {\hbar^2 \over
2} \int_{{\bf R}^2} \sigma \big(\tilde{A} \buildrel{\bullet}\over{\wedge}
\star \tilde{A} \big) dp \wedge dq, $$

$$ = - {\hbar^2 \over 2} \int_{{\bf R}^2} \sigma\big(g^{-
\buildrel{\bullet}\over{1}} \bullet dg \buildrel{\bullet}\over{\wedge}
\star g^{- \buildrel{\bullet}\over{1}}\bullet dg \big) dp \wedge dq.
\eqno(6) $$ This Lagrangian has precisely equation of motion (5). One can
apply also the above procedure to the well known Yang and
Donaldson-Nair-Schiff equations. In Ref. 1 we found
that these equations admit suitable Moyal deformation via Fedosov's
geometry. One can follow the same procedure in order to reveal the
underlying geometry of deformation quantization of the Moyal deformed
WZW-like action of SDG obtained in Ref. 2.

\section{Final Remarks}

Some further questions remain to be overcome. For instance, it would be
very interesting to investigate the behavior of heavenly hierarchies of
conserved quantities within Fedosov's geometry and some other alternative
geometries of deformation quantization.  There exists a strong relation
between SDYM and SDG with the M(atrix) theory approach to $M$-theory
\cite{Banks}. A similar relation exists between the later and $N=(2,1)$
strings\cite{Martinec}. $N=(2,1)$ strings are also related to SDG through
$N=2$ heterotic strings.  Since in both descriptions of $M$-theory is
involved SDYM and SDG one would hope that some new geometrical
descriptions of SDYM theory and SDG will be of some relevance to give more
insight into $M$-theory.

\end{document}